\documentclass{iopart}

\def\bea{\begin{eqnarray}}
\def\ena{\end{eqnarray}}

\def\spacing{\baselineskip=20pt}

\usepackage{graphicx}
\usepackage{dcolumn}
\usepackage{amssymb}

\begin{document}

\title{Components of the gravitational force in the field of a gravitational wave}

\author{D. Baskaran\dag\ and L. P. Grishchuk\dag\ddag
\footnote[3]{e-mails: spxdb@astro.cf.ac.uk,
grishchuk@astro.cf.ac.uk} }

\address{\dag\ Department of Physics
and Astronomy, Cardiff University, Cardiff CF24 3YB, United
Kingdom}

\address{\ddag\ Sternberg Astronomical Institute, Moscow University, Moscow 119899, Russia}


\begin{abstract}

Gravitational waves bring about the relative motion of free test
masses. The detailed knowledge of this motion is important
conceptually and practically, because the mirrors of laser
interferometric detectors of gravitational waves are essentially
free test masses. There exists an analogy between the motion of
free masses in the field of a gravitational wave and the motion of
free charges in the field of an electromagnetic wave. In
particular, a gravitational wave drives the masses in the plane of
the wave-front and also, to a smaller extent, back and forth in
the direction of the wave's propagation. To describe this motion,
we introduce the notion of ``electric" and ``magnetic" components
of the gravitational force. This analogy is not perfect, but it
reflects some important features of the phenomenon. Using
different methods, we demonstrate the presence and importance of
what we call the ``magnetic" component of motion of free masses.
It contributes to the variation of distance between a pair of
particles. We explicitly derive the full response function of a
2-arm laser interferometer to a gravitational wave of arbitrary
polarization. We give a convenient description of the response
function in terms of the spin-weighted spherical harmonics. We
show that the previously ignored ``magnetic" component may provide
a correction of up to 10$\%$, or so, to the usual ``electric"
component of the response function. The ``magnetic" contribution
must be taken into account in the data analysis, if the parameters
of the radiating system are not to be mis-estimated.

\end{abstract}


\spacing


\section{\label{sec:introduction}Introduction}

The large-scale interferometers for the detection of gravitational
waves are approaching their planned level of sensitivity
\cite{ligo}. In the coming years, we may be observing the rare,
but most powerful, sources of gravitational waves (for reviews,
see for example [2-7]). These initial instruments are scheduled to
be upgraded in a few years time. The upgraded (advanced)
interferometers will be more sensitive than the initial ones. They
will enable us to determine with high accuracy the physical
parameters of various astrophysical sources. The knowledge of the
detailed and accurate response function of the instrument to the
incoming gravitational wave (g.w.) of arbitrary polarization
becomes an important priority.

A laser interferometer monitors distances between the central
mirror(s) and the end-mirrors in the interferometer's arms. In the
existing instruments, the length $l$ of the interferometer's arm
is significantly shorter than the wavelengths $\lambda$ of the
gravitational waves which the instrument is most sensitive to. The
evaluation of the change $\delta l$ of the distance between two
mirrors is usually formulated as $\delta l/l  \approx h$, where
$h$ is the characteristic amplitude of the incoming wave. This is
a correct answer, but it is correct only in the lowest,
zero-order, approximation in terms of the small parameter $l/
\lambda$. The next approximation introduces a ``magnetic"
correction \cite{gr77} to $\delta l/l$, which is of the order of
$h (l/\lambda)$. This correction depends on the ratio $l/\lambda$
whose numerical value is at the level of several percents for the
instruments with $l = 4km$ and typical g.w. frequencies $\nu =
c/\lambda \approx 10^{3} Hz$. This contribution may be especially
important for the advanced interferometers operating in the
``narrow-band" mode aimed at detecting relatively high-frequency
quasi-monochromatic g.w. signals.

In this paper, we start from the motion of free charges in the
field of an electromagnetic wave,
section~\ref{sec:electrodynamics}. Then, we turn our attention to
the main subject of the paper -- motion of free masses in the
field of a gravitational wave. In section~\ref{sec:inertialframe}
the positions and motion of free test masses is described in the
local inertial reference frame associated with one of the masses.
This choice of coordinates is the closest thing to the global
Lorentzian coordinates that are normally used in electrodynamics.
The identification of the ``electric" and ``magnetic" components
of motion, as well as comparison with electrodynamics, are
especially transparent in this description. In
section~\ref{sec:geodesicdeviation} we exploit the geodesic
deviation equation for the derivation of equations of motion and
identification of the components of the  gravitational force. The
usually written equation, with the curvature tensor in it, is only
the zero-order approximation in terms of $l/\lambda$. This
approximation is sufficient for the description of the ``electric"
part of the motion, but is insufficient for the description of the
``magnetic" part. In the next approximation, which is a first
order in terms of $l/\lambda$, the geodesic deviation equation
includes the derivatives of the curvature tensor, and this
approximation is fully sufficient for the description of the
``magnetic" force and ``magnetic" component of motion. From these
considerations it follows that the component of motion which we
call, with some reservations, ``magnetic" is, in any case, the
finite-wavelength correction to the usual infinite-wavelength
approximation. In the end of this section we compare our
interpretation with other analogies cited in the literature. In
section~\ref{sec:distances} we switch from the positions and
trajectories of individual particles to the distances between
them. The calculation of distances is especially simple in the
local inertial frame, since the metric tensor, in the required
approximation, is simply the Minkowski tensor. However, the
conclusions about distances, including their ``magnetic"
contributions, do not depend on the choice of coordinates. We show
that the universal definitions of distance, based on the light
travel time and the length of spatial geodesics, lead to the same
result in the appropriate approximation. In
section~\ref{sec:interferometer} we use the derived results for
the construction of the response function of a 2-arm ground-based
interferometer. It is assumed that a gravitational wave of
arbitrary polarization is coming from arbitrary direction on the
sky. We explicitly identify the ``magnetic" contribution to the
response function and demonstrate its importance. Then in
section~\ref{sec:spinfunction} we formulate the response function
in terms of the spin-weighted spherical harmonics. In
section~\ref{sec:discussion} we give numerical estimates of the
``magnetic" contribution to the response function of the presently
operating instruments, in the context of realistic astrophysical
sources. We show that the lack of attention to the ``magnetic"
components can lead to a considerable systematic error in the
estimation of parameters of the g.w. source. Some technical
details are described in Appendices.


\section{\label{sec:electrodynamics}Motion of free charges
in the field of electromagnetic wave}

To understand the origin of the gravitational ``magnetic"
contribution to the motion of free masses, it is convenient to
start the analysis with the similar problem in electrodynamics.
Let us consider the motion of a charged particle in the field of a
monochromatic electromagnetic wave. It is known \cite{ll} that a
charged particle in the electromagnetic field is subject to the
electromagnetic Lorentz force $\textbf{F}$ given by \bea
\textbf{F} =
e\textbf{E}+\frac{e}{c}\left(\textbf{v}\times\textbf{H}\right).
\label{lorentz} \ena The first term in Eq. (\ref{lorentz}) is the
electric contribution to the force, while the second term is the
magnetic contribution. The ratio of the second term to the first
one is of the order of $v/c$. This means that in the field of a
weak electromagnetic wave, i.e. in the field of a wave that gives
rise to a small velocity $v/c \ll 1$ of the charged particle, the
magnetic contribution to the force $F$ is also small. To find the
trajectory of the particle, one has to solve the equations \bea
m\frac{d^2\textbf{x}}{dt^2}=e\textbf{E}+\frac{e}{c}\left(\frac{d\textbf{x}}{dt}
\times\textbf{H}\right). \label{newtonelectrodynamics} \ena The
character of the trajectory depends on the wave's polarization.

\begin{figure}
\begin{center}
\includegraphics{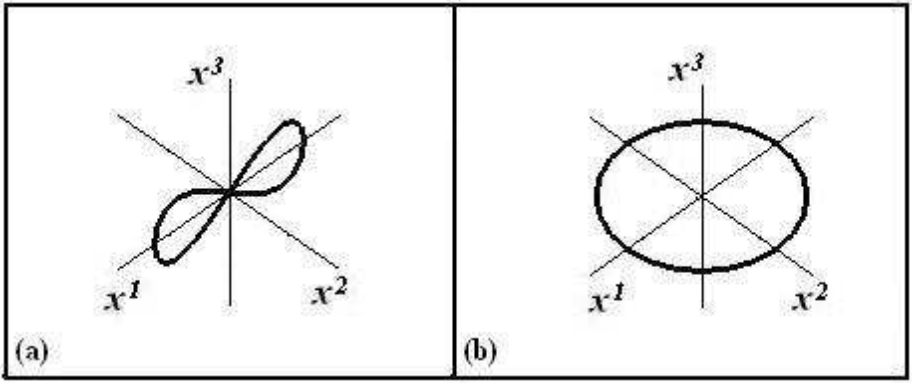}
\end{center} \caption{\label{fig:electromagneticcase}
The figure a) on the left shows the trajectory of a charged
particle in the field of a linearly polarized electromagnetic
wave, whereas the figure b) on the right shows the trajectory in
the field of a circularly polarized wave.} \end{figure}

Let us start with a linearly polarized monochromatic wave of
angular frequency $\omega$, propagating in the $x^3$ direction,
with the electric field $E$ oscillating along the $x^1$ axis. The
trajectory of the particle, which is on average at rest, is given
by \cite{ll} \bea \begin{array}{c} x^1 = -
\frac{eE}{m\omega^{2}}\cos{\omega t},\ \ \ x^2 = 0, \ \ \ x^3 =
\frac{eE}{8m\omega^{2}}\left(\frac{eE}{mc\omega}\right)
\sin{2\omega t}.
\end{array}
\ena The particle moves in the $(x^1, x^3)$ plane along a
symmetric curve, shaped as figure 8, with its main axis oriented
in the $x^1$ direction, see Fig.~\ref{fig:electromagneticcase}a.
The electric component of the Lorentz force is dominant, and it
drives the particle in the plane of the wave-front, while the
magnetic component is responsible for the movement of the particle
back and forth in the $x^3$ direction. For a weak electromagnetic
wave, the size of the $x^1$-amplitude is small in comparison with
the wavelength: $\frac{x^1}{\lambda} \sim \frac{eE}{mc\omega}\ll
1$, but the size of the $x^3$-amplitude is even smaller: $x^3 \sim
x^1 \left( \frac{x^1}{\lambda} \right)$. The ratio of $x^3$ and
$x^1$ amplitudes is of the order of $\frac{eE}{mc\omega}\sim
\frac{v}{c}$, where $v$ is the characteristic value of the
particle's velocity.

A circularly polarized wave provides an exceptional case wherein
the magnetic component of the force is inactive; the particle does
not have an $x^3$-component of motion. The trajectory of the
particle, which is on average at rest, is given by \cite{ll}, \bea
\begin{array}{c} x^1 = - \frac{eE}{m\omega^{2}}\cos{\omega t},\ \
\ x^2 =-\frac{eE}{m\omega^{2}}\sin{\omega t}, \ \ \ x^3 = 0.
\end{array} \ena The particle moves in the $(x^1, x^2)$ plane
along a circle, as shown in Fig.~\ref{fig:electromagneticcase}b.
In the general case of an elliptically-polarized wave, the
$x^3$-component of motion is always present. The projection of the
trajectory onto the $(x^1, x^2)$ plane forms an ellipse. The
ellipse degenerates into a straight line or a circle for a
linearly-polarized or a circularly-polarized wave, respectively.
We shall see below that there exists an analogy between the
considered motion of charged particles in the field of an
electromagnetic wave and the motion of test masses in the field of
a gravitational wave.


\section{\label{sec:inertialframe}Motion of a free test mass
in the field of a weak gravitational wave}

Weak gravitational waves belong to the class of weak gravitational
fields: \bea ds^{2}= c^{2}dt^{2} - \left[ \delta_{ij} + h_{ij}
\right] dx^{i}dx^{j}. \label{metricgmunu} \ena A general
expression for a plane wave incoming from the positive $x^3$
direction is given by \bea h_{ij} = {\stackrel{1}{p}_{ij}} a +
{\stackrel{2}{p}_{ij}} b, \label{sol} \ena where \bea
\begin{array}{l} \label{abg} a =
h_{+}\sin{\left(k({x^0}+{x^3})+\psi_{+}\right)}, \ \ \ b =
h_{\times}\sin{\left(k({x^0}+{x^3})+\psi_{\times}\right)},
\end{array} \ena and $k=\frac{2\pi}{\lambda}=\frac{2\pi\nu}{c}=\frac{\omega}{c}$,
$x^0 =ct$; $\psi_{+}$ and $\psi_{\times}$ are arbitrary phases.
The polarization tensors ${\stackrel{s}{p}_{ij}}$ $(s=1,2)$ are
given by \bea \begin{array}{l} \label{pol}
{\stackrel{1}{p}_{ij}}=m_{i}m_{j}-n_{j}n_{i},\ \ \ \
{\stackrel{2}{p}_{ij}}= -m_{i}n_{j}-m_{j}n_{i}, \end{array} \ena
where the unit vectors $m_{i}$ and $n_{i}$, together with the unit
vector $k_3/k$ pointing out in the direction of the wave
propagation, form a triplet of mutually orthogonal unit vectors.

One still has the freedom of turning the coordinate system in the
($x^1, x^2$) plane by some angle. This transformation mixes the
components of the matrix (\ref{sol}). Using this freedom (see
\ref{app:coordtransf1}) one can simplify the functions (\ref{abg})
in such a way that in the new coordinate system they take the form
\bea \begin{array}{l} a =
h_{+}\sin{\left(k({x^0}+{x^3})+\psi\right)}, \ \ \ b =
h_{\times}\cos{\left(k({x^0}+{x^3})+\psi\right)},
\end{array}
\label{ab} \ena where $h_{+}, ~h_{\times}$ are arbitrary
amplitudes and $\psi$ is an arbitrary phase. The unit vectors
$m_{i}, ~n_{i}$ have the components $m_i = 1,0,0$ and $n_i=
0,1,0$. We shall call this special coordinate system a frame based
on principal axes. Two independent linear polarization states are
defined by the conditions $h_{+}=0$ or $h_{\times}=0$, and two
independent states of circular polarization are defined by
$h_{+}=h_{\times}$ or $h_{+}= -h_{\times}$. Specifically, we will
be working with the metric \bea ds^2
=c^2dt^2-(1+a)d{x^1}^2-(1-a)d{x^2}^2+2b~d{x^1}d{x^2}-d{x^3}^{2},
\label{metric} \ena where the functions $a$ and $b$ are given by
Eq. (\ref{ab}).

The metric (\ref{metric}) is written in synchronous coordinates.
Therefore, the world lines ${x^1},{x^2},{x^3}=const$ are time-like
geodesics, they represent the histories of free test masses. To be
as close as possible to the framework of laboratory physics, and
specifically to electrodynamic examples considered in the previous
section, we have to introduce a local inertial coordinate system
$(\bar{t},{\bar{x}^1},{\bar{x}^2},{\bar{x}^3})$. Let us associate
it with the central world line ${x^1}={x^2}={x^3}=0$. By
definition, in the local inertial frame, and along the central
geodesic line, the metric tensor takes on the Minkowski values and
all first derivatives of the metric tensor vanish. The transformed
metric takes on the form \bea \bar{g}_{\mu\nu}= \eta_{\mu\nu}+
{\it terms~of~the~order~of}~\left(h\frac{{\overline{x}^1}^2
+{\overline{x}^2}^2+{\overline{x}^3}^2}{\lambda^{2}}\right).
\label{limetr} \ena A local inertial frame realizes, as good as we
can do in the presence of the gravitational field, a rigid freely
falling ``box" with a clock in it \cite{mtw}. The required
coordinate transformation
${\bar{x}}^{\mu}={\bar{x}}^{\mu}\left(x^{\nu}\right)$ is given by
\cite{gr77}: \bea
\begin{array}{l}
{\bar{x}^0} = x^0 + \frac{1}{4}\dot{a}~
\left({x^1}^2-{x^2}^2\right)-\frac{1}{2}\dot{b}~
{x^1}{x^2}, \\ \\
{\bar{x}^1} = {x^1} +\frac{1}{2}a~ {x^1}-\frac{1}{2}b~
{x^2}+\frac{1}{2}\dot{a}~ {x^3}{x^1} -\frac{1}{2}\dot{b}~
{x^3}{x^2}, \\ \\ {\bar{x}^2} = {x^2} -\frac{1}{2}a~
{x^2}-\frac{1}{2}b~ {x^1}-\frac{1}{2}\dot{a}~ {x^3}{x^2}
-\frac{1}{2}\dot{b}~ {x^3}{x^1}, \\ \\ {\bar{x}^3} = {x^3} -
\frac{1}{4}\dot{a}~
\left({x^1}^2-{x^2}^2\right)+\frac{1}{2}\dot{b}~ {x^1}{x^2}.
\end{array} \label{barxyz1} \ena In the above transformation, the
functions $a$, $b$ and their time derivatives
$\dot{a}=\frac{1}{c}\frac{\partial a}{\partial t}$,
$\dot{b}=\frac{1}{c}\frac{\partial b}{\partial t}$ are evaluated
along the world line ${x^1}={x^2}={x^3}=0$. The linear and
quadratic terms, as powers of $x^{\alpha}$, are unambiguously
determined by the conditions of local inertial frame, but the
cubic and higher-order corrections are not determined by these
conditions. In principle, transformations (\ref{barxyz1}) can be
used for all values of $x^{\alpha}$, but they are physically
useful when the values of $x^{\alpha}$ are sufficiently small,
that is, when the cubic and higher-order terms can be neglected.

Let us consider a free test mass riding on a time-like geodesic
(${x^1=l_{1}}$, ${x^2=l_{2}}$, ${x^3=l_{3}}$). Equations
(\ref{barxyz1}) define the behaviour of this mass with respect to
the introduced local inertial frame. Concretely, we have \bea \fl
\begin{array}{l} {\bar{x}^1}(t) = {l_{1}}
+\frac{1}{2}\left[h_{+}{l_{1}}\sin{(\omega
t+\psi)}-h_{\times}{l_{2}}\cos{(\omega t+\psi)}\right]\\ \\
~ \ \qquad + \frac{1}{2}k\left[h_{+}{l_{3}}{l_{1}}\cos{(\omega
t+\psi)}
+h_{\times}{l_{3}}{l_{2}}\sin{(\omega t+\psi)}\right], \\ \\
{\bar{x}^2}(t) = {l_{2}}
-\frac{1}{2}\left[h_{+}{l_{2}}\sin{(\omega
t+\psi)}+h_{\times}{l_{1}}\cos{(\omega t+\psi)}\right]\\ \\
~ \ \qquad - \frac{1}{2}k\left[h_{+}{l_{3}}{l_{2}}\cos{(\omega
t+\psi)}
-h_{\times}{l_{3}}{l_{1}}\sin{(\omega t+\psi)}\right], \\ \\
{\bar{x}^3}(t) = {l_{3}} -
\frac{1}{4}k\left[h_{+}\left({l_{1}}^2-{l_{2}}^2\right)\cos{(\omega
t+\psi)}+ 2h_{\times}{l_{1}}{l_{2}}\sin{(\omega t+\psi)}\right].
\end{array}
 \label{barxyz}
\ena Obviously, the unperturbed ({\it i.e.} in the absence of the
gravitational wave) position of the nearby  mass is
$({l_{1}},{l_{2}},{l_{3}})$. The action of the wave drives the
mass in an oscillatory fashion around the unperturbed position. In
general, all three components of motion are present.

\begin{figure}
\begin{center}
\includegraphics{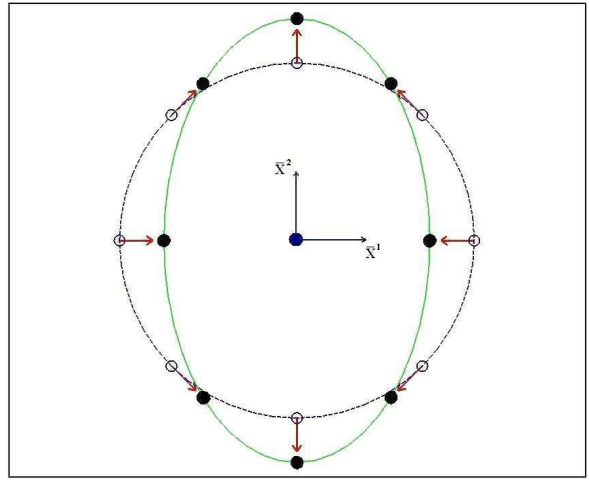}
\end{center} \caption{\label{fig:tensor}
Motion of free test masses in the field of a linearly polarized
($h_{+}\neq 0, h_{\times}=0$) g.w. in the lowest approximation,
that is, when the wave number $k$ is effectively set to zero.}
\end{figure}

If one neglects in Eq. (\ref{barxyz}) the terms with $k$, thus
effectively sending the wavelength $\lambda$ to infinity, one
arrives at the often cited statement about the particle's motion:
\bea \begin{array}{l} {\bar{x}^1}(t) = {l_{1}}
+\frac{1}{2}\left[h_{+}{l_{1}}\sin{(\omega
t+\psi)}-h_{\times}{l_{2}}\cos{(\omega t+\psi)}\right],\\ \\
{\bar{x}^2}(t) = {l_{2}}
-\frac{1}{2}\left[h_{+}{l_{2}}\sin{(\omega
t+\psi)}+h_{\times}{l_{1}}\cos{(\omega t+\psi)}\right],\\ \\
{\bar{x}^3}(t) = {l_{3}}.
\end{array}
\label{barxyzE} \ena Clearly, this is the analog of the electric
component of motion in electrodynamics; moving particles remain in
the plane of the wave-front. The oscillations of individual
particles, for the particular case of a linearly polarized g.w.,
are shown in Fig.~\ref{fig:tensor}.

\begin{figure}
\begin{center}
\includegraphics{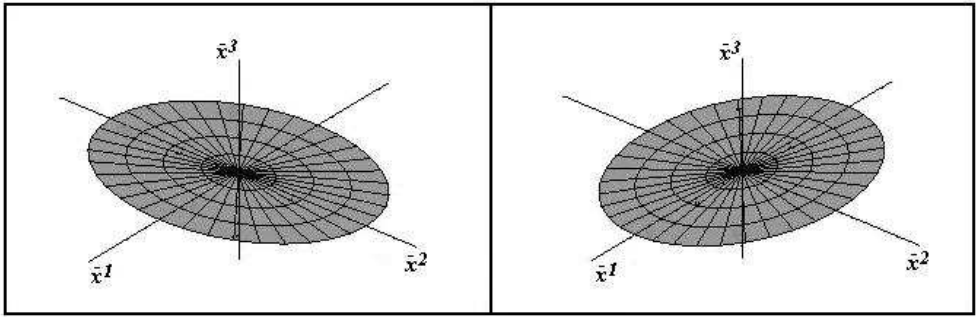}
\end{center} \caption{\label{fig:electric}
Deformation of a disk of free test particles in the field of a
linearly polarized ($h_{+}\neq 0, h_{\times}=0$) g.w. in the limit
of $k=0$. The two figures show the displacements at the moments of
time separated by a half period.} \end{figure}

A circular disk, consisting of particles that are located in the
plane of the wave-front, stretches and squeezes in an oscillatory
fashion, as shown in Fig.\ref{fig:electric}.

\begin{figure}
\begin{center}
\includegraphics{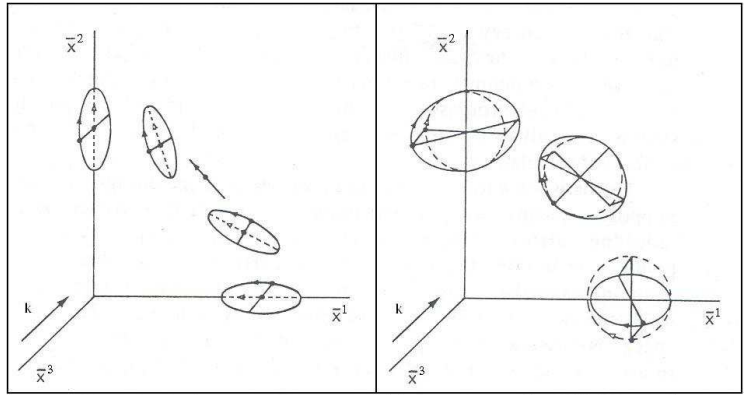}  \end{center} \caption{\label{fig:movement}
Solid lines show 3-dimensional motion of particles, while dashed
lines show projections of the trajectories onto the
$\bar{x}^{1},\bar{x}^{2}$ plane. In the left picture the wave is
linearly polarized ($h_{+}\neq 0, h_{\times}=0$), whereas in the
right picture the wave is circularly polarized
($h_{+}=h_{\times}$).} \end{figure}

Now, let us take into account the terms with $k$ in Eq.
(\ref{barxyz}). If $l_3 =0$, these terms do not change the
${\bar{x}^1},{\bar{x}^2}$ components of motion, but nevertheless
they introduce oscillations along the ${\bar{x}^3}$ direction. In
analogy with electrodynamics, it is reasonable to call these terms
the ``magnetic" components of motion. The trajectories of free
masses are, in general, ellipses, and they are not confined to the
plane of the wave-front. In Fig.~\ref{fig:movement} (taken from
\cite{gr77}) we show the trajectories of some individual
particles. The ``magnetic" contribution is smaller than the
``electric" one. In analogy with electrodynamics, the major axis
of the individual ellipse is small in comparison with $\lambda$
and $l$, but the size of the ${\bar{x}^3}$-amplitude is even
smaller than the ${\bar{x}^1}$-amplitude. Their ratio is typically
of the order of $kl=2\pi (l/\lambda)$, where
$l=\sqrt{{l_1}^2+{l_2}^2+{l_3}^2}$ is the mean (unperturbed)
distance of the test particle from the origin. In analogy with
electrodynamics, the ``magnetic" component of motion will be
present even if the particle is initially at rest with respect to
the local inertial frame.

\begin{figure}
\begin{center}
\includegraphics{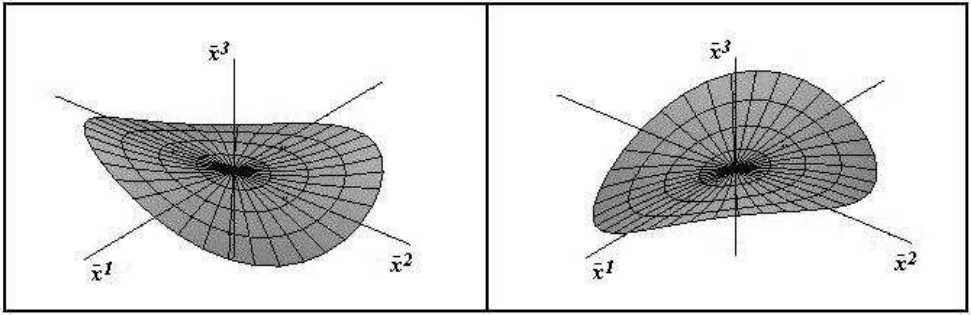}  \end{center} \caption{\label{fig:magnetic}
The figure shows the deformations of a circular disk of free test
particles under the action of a linearly polarized g.w.
($h_{+}\neq 0, h_{\times}=0$). The ``magnetic" contribution is
responsible for the displacements along the ${\bar{x}^3}$ axis.
The two pictures show the configurations at the moments of time
separated by a half of period.} \end{figure}

A circular disk, consisting of free test particles, also behaves
differently, as compared with the lowest order ``electric"
approximation. The disk does not remain flat while being stretched
and squeezed. In addition to being stretched and squeezed it
bends, in an oscillatory manner, forward and backward in the
${\bar{x}^3}$ direction. We show these complicated deformations in
Fig.~\ref{fig:magnetic}. The level of the ``magnetic" contribution
is determined by the assumption that $l/\lambda = 0.1$. This
figure should be compared with Fig.~\ref{fig:electric}.

\section{\label{sec:geodesicdeviation} Equations of motion from the geodesic deviation equations.}

In this section we shall consider the motion of free masses from a
different perspective. The geodesic deviation equations will be
used in order to derive the analog of the Lorentz force Eq.
(\ref{lorentz}) and the analog of the Newtonian equations of
motion Eq. (\ref{newtonelectrodynamics}). We shall see that the
``magnetic" component of motion is contained in the higher-order
geodesic deviation equations.


\subsection{\label{deviationsubsection1}Geodesic deviation equations in general}

\begin{figure}
\begin{center}
\includegraphics{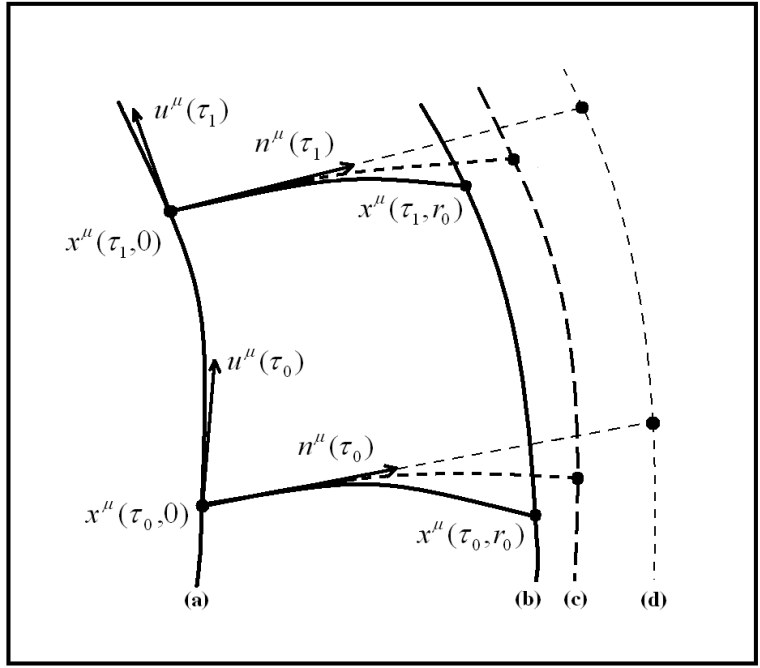}  \end{center}
\caption{\label{fig:geodesics} Deviation of two nearby geodesics
in a gravitational field. Line (a) represents the central geodesic
$r=0$ and line (b) represents the nearby geodesic $r=r_{0}$.
$u^{\mu}$ is the unit tangent vector to the central world line,
$n^{\mu}$ is the tangent vector to the curve $\tau =const$. The
lines (d) and (c) are the corresponding first and second order
approximations to the exact nearby geodesic (b).} \end{figure}

The derivation of the geodesic deviation equations is usually
based on a 2-parameter family of time-like geodesics
$x^{\mu}(\tau,r)$. For each value of $r$, the line
$x^{\mu}(\tau,r)$ is a time-like geodesic with a proper time
parameter $\tau$. The vector $u^{\mu}$ is the unit tangent vector
to the geodesic, and $n^{\mu}$ is the ``separation" vector between
the geodesics (see Fig.~\ref{fig:geodesics}): \bea
u^{\mu}(\tau,r)=\frac{\partial x^{\mu}}{\partial
\tau}|_{r=const}~, ~ ~ ~ n^{\mu}(\tau,r)=\frac{\partial
x^{\mu}}{\partial r}|_{\tau =const}~. \ena

Let the central (fiducial) geodesic line correspond to $r=0$, and
a nearby geodesic to $r=r_0$. For small $r_0$, one has: \bea
x^{\mu}(\tau,r_0) = x^{\mu}(\tau,0) + r_0\frac{\partial
x^{\mu}}{\partial r} + \frac{1}{2} r_0^2 \frac{\partial^2
x^{\mu}}{{\partial r}^2} + O(r_0^3). \label{expan} \ena Then, in
the lowest approximation in terms of $r_0$, the geodesic deviation
equations are given by \cite{mtw}: \bea
\frac{D^{2}n^{\mu}}{d\tau^{2}}=R^{\mu}_{\alpha \beta
\gamma}u^{\alpha}u^{\beta}n^{\gamma}, \label{deviation1} \ena
where $R^{\mu}_{\alpha \beta \gamma}$ is the curvature tensor
calculated along the geodesic line $r=0$, and $\frac{D}{d \tau}$
is the covariant derivative calculated along that line.

To discuss the ``magnetic" component of motion in the field of a
gravitational wave we will need the geodesic deviation equations
extended to the next approximation. These equations were derived
by Bazanski \cite{Bazanski}. A modified derivation can be found in
\cite{Kerner}. In the required approximation, one needs the
information on the second derivatives of $x^{\mu}$:
$\frac{\partial^2 x^{\mu}}{{\partial r}^2}$. These quantities do
not form a vector. It is convenient to introduce a closely related
vector $w^{\mu}$: \bea w^{\mu}=\frac{Dn^{\mu}}{dr} =
n^{\mu}_{;\alpha}n^{\alpha} = \frac{\partial^2 x^{\mu}}{{\partial
r}^2}+ \Gamma^{\mu}_{\alpha \beta} n^{\alpha} n^{\beta}.
\label{vecw} \ena This vector obeys the equations \cite{Bazanski}
(see also \cite{Kerner}): \bea \fl \frac{D^{2}w^{\mu}}{d\tau^{2}}=
R^{\mu}_{\alpha \beta \gamma}u^{\alpha}u^{\beta}w^{\gamma}
+\left(R^{\mu}_{\alpha \beta \gamma ; \delta}-R^{\mu}_{\gamma
\delta \alpha ;
\beta}\right)u^{\alpha}u^{\beta}n^{\gamma}n^{\delta}
+4R^{\mu}_{\alpha \beta
\gamma}u^{\beta}\frac{Dn^{\alpha}}{d\tau}n^{\gamma}.
\label{deviation2} \ena

To combine equations (\ref{deviation1}) and (\ref{deviation2})
into a single formula, valid up to and including the terms of the
order of $r_0^2$, it is convenient to construct a vector
$N^{\mu}$: \bea N^{\mu}=r_{0}  n^{\mu}+\frac{1}{2}r_{0}^{2}
w^{\mu}. \ena Taking the sum of Eq. (\ref{deviation1}), multiplied
by $r_0$, and Eq. (\ref{deviation2}) multiplied by $r_0^2/2$, we
obtain the equation whose solution we will need,  \bea\fl
\frac{D^{2}N^{\mu}}{d\tau^{2}}= & R^{\mu}_{\alpha \beta
\gamma}u^{\alpha}u^{\beta}N^{\gamma}
 +\frac{1}{2}\left(R^{\mu}_{\alpha \beta \gamma ;
\delta}-R^{\mu}_{\gamma \delta \alpha ;
\beta}\right)u^{\alpha}u^{\beta}N^{\gamma}N^{\delta} \nonumber \\
\fl & + 2R^{\mu}_{\alpha \beta
\gamma}u^{\beta}\frac{DN^{\alpha}}{d\tau} N^{\gamma} + O(r_{0}^3).
\label{deviation3} \ena In terms of $N^{\mu}$, the expansion
(\ref{expan}) takes the form \bea x^{\mu}(\tau,r_0) =
x^{\mu}(\tau,0) + N^{\mu} - \Gamma^{\mu}_{\alpha \beta} N^{\alpha}
N^{\beta} +O(r_0^3). \label{expan2} \ena Formula (\ref{expan2})
shows that in the local inertial frame (in which case
${\Gamma}^{\mu}_{\alpha\beta}=0$ along the central geodesic line)
the spatial components $N^i$ will directly give the time-dependent
positions of the nearby particle. According to Eq.
(\ref{deviation3}), these positions include the next order
corrections, as compared with solutions to Eq. (\ref{deviation1}).

In Fig~\ref{fig:geodesics} we show the successive approximations
to the exact position of the nearby geodesic $(b)$. The line $(d)$
represents the first order approximation ({\it i.e.}
$x^{\mu}(\tau,r_{0})=x^{\mu}(\tau,0)+r_{0}\frac{\partial
x^{\mu}}{\partial r}$ ), the line $(c)$ takes into account the
second order approximation ({\it i.e. $x^{\mu}(\tau,r_0) =
x^{\mu}(\tau,0) + r_0\frac{\partial x^{\mu}}{\partial r} +
\frac{1}{2} r_0^2 \frac{\partial^2 x^{\mu}}{{\partial r}^2}$} ),
according to (\ref{expan}).

\subsection{\label{deviationsubsection3}Geodesic deviation in the field of a gravitational wave}

We now specialize to the g.w. metric (\ref{metricgmunu}),
(\ref{sol}), where $a$ and $b$ are given by (\ref{ab}). We take
account only of linear perturbations in terms of g.w. amplitude
$h$. The first particle is described by the central time-like
geodesic $x^i(t)=0$, its tangent vector is $u^{\alpha} =
(1,0,0,0)$. The second particle resides at the unperturbed
position $x^i(0)=l^i$ and has zero unperturbed velocity. It is
assumed that the local inertial frame is realized along the
central geodesic. The objective is to find the trajectory of the
second particle using the geodesic deviation equation
(\ref{deviation3}).

The deviation vector $N^i$ has the form \bea
{N}^{i}(t)=l^{i}+\xi^{i}(t), \ena where $\xi^{i}(t)$ is the
perturbation caused by the gravitational wave. The choice of the
local inertial frame allows one to replace all covariant
derivatives in Eq. (\ref{deviation3}) by ordinary derivatives. In
particular, covariant time derivative $\frac{D^2}{d\tau^2}$ is
being replaced by ordinary time derivative $\frac{d^2}{c^2dt^2}$.
In the lowest approximation, Eqs. (\ref{deviation3}) reduce to
Eqs. (\ref{deviation1}) and specialize to \bea \fl
\frac{d^{2}\xi^{i}}{dt^{2}}= - \frac{1}{2}l^{j}\frac{\partial^{2}}
{{\partial t}^{2}}\left({h^i_j}\right)  =
\frac{1}{2}\omega^{2}l^{j} \left[ {\stackrel{1}{p}{}_{j}^{i}}h_{+}
\sin(\omega t+\psi)+{\stackrel{2}{p}{}_{j}^{i}}h_{\times}
\cos(\omega t+\psi) \right]. \label{eqmel}\ena As expected, the
relevant solution to this equation coincides exactly with the
usual ``electric" part of the motion, which is given by Eq.
(\ref{barxyzE}).

In order to identify the ``magnetic" part of the gravitational
force, one has to consider all the terms in Eq.
(\ref{deviation3}). Since $DN^{\alpha}/d\tau$ is of the order of
$h$, the third term in Eq. (\ref{deviation3}) is of the order of
$h^2$ and should be neglected. Working out the derivatives of the
curvature tensor and substituting them into (\ref{deviation3}), we
arrive at the accurate equations of motion: \bea \fl
\frac{d^{2}\xi^{i}}{dt^{2}} = - \frac{1}{2}l^j \frac{\partial^{2}}
{{\partial t}^{2}}\left(h^i_j\right) - \frac{1}{2}l^k l^l
\frac{\partial^{2}}{{\partial t}^{2}}
\left(\frac{\partial}{\partial x^l}\left(h^i_k\right) -\frac{1}{2}
\delta^{ij} \frac{\partial}{\partial x^j}\left( h_{kl}\right)
\right). \label{eqmfull}\ena The second term of this formula is
responsible for the ``magnetic" component of motion and can be
interpreted as the gravitational analog of the magnetic part of
the Lorentz force (\ref{lorentz}).

Specifically, in the field of the gravitational wave
(\ref{metric}), the full equations of motion (\ref{eqmfull}) take
the form:  \bea \fl \frac{d^{2}\xi^{i}}{dt^{2}} =&
\frac{1}{2}\omega^{2}l^{j}\left[ {\stackrel{1}{p}{}_{j}^{i}}h_{+}
\sin(\omega t+\psi)+{\stackrel{2}{p}{}_{j}^{i}}h_{\times}
\cos(\omega t+\psi)\right] \nonumber\\ \fl & - \frac{1}{2}
\omega^2 l^k l^l \left[k_l \delta^{ij} +\frac{1}{2}k^i \delta^j_l
\right] \left[{\stackrel{1}{p}{}_{kj}}h_{+} \cos(\omega
t+\psi)-{\stackrel{2}{p}{}_{kj}}h_{\times} \sin(\omega
t+\psi)\right]. \label{eqmfull2}\ena This equation clearly
exhibits two contributions: \bea
m\frac{d^2\xi^i}{dt^2}=F_{(e)}^i+F_{(m)}^i.\ena The ``electric"
component of the gravitational force $F_{(e)}^i/m$ is given by the
first term in Eq. (\ref{eqmfull2}). The second term - the
``magnetic" component $F_{(m)}^i/m$ of the gravitational force -
can be written, demonstrating certain analogy with
electrodynamics, in the form involving the (lowest order
non-vanishing) velocity $\frac{d\xi^i}{dt}$ of the test particle:
\bea \frac{F_{(m)}^i}{m} =   \omega l^l\left[k_l \delta^{i}_j
+\frac{1}{2}k^i \delta_{jl} \right]\frac{d\xi^j}{dt}.\ena

The right hand side of Eq. (\ref{eqmfull2}) is exactly the
acceleration $\frac{d^2\bar{x}^i}{dt^2}$ which can be derived by
taking the time derivatives of Eq. (\ref{barxyz}). Not
surprisingly, by integrating equations of motion (\ref{eqmfull2}),
one arrives exactly at the time-dependent positions of the
particles, Eq. (\ref{barxyz}), which we have already derived by
the direct coordinate transformation. Therefore, the gravitational
Lorentz force, identified above, leads exactly to the expected
result.

It should be noted that Eqs. (\ref{eqmfull2}) depend only on the
so-called TT-components of the g.w. field. This happens not only
because we have explicitly started from them in Eq.
(\ref{metric}). Even if we have started from the general form of
the g.w. field, which includes also the non-TT components, we
would have ended up with equations containing only the
TT-components. This happens because Eqs. (\ref{deviation3})
involve the curvature tensor (and its derivatives) in which the
non-TT components automatically cancel out.

It is interesting to compare the components of the gravitational
force derived here with what would follow from the concept of
``gravitomagnetism". In general, the concept of
``gravitomagnetism" is a helpful analogy which was successfully
used in studies of stationary gravitational fields \cite{wald},
\cite{mashhoon}. However, its application to the
gravitational-wave problem considered here requires certain care.
For example, in the local inertial frame, the leading terms of the
equations of motion derived in the framework of
``gravitomagnetism" \cite{mashhoon} read (in notations consistent
with this paper):
 \bea\fl
\frac{d^{2}N^{i}}{dt^{2}}= R^{i}_{0 0 j}N^{j}
  + 2R^{i}_{j 0
k}\frac{dN^j}{dt} N^{k}. \label{deviationmashhoon} \ena This
equation should be compared with our Eq. (\ref{deviation3}), also
specialized to the local inertial frame: \bea\fl
\frac{d^{2}N^{i}}{dt^{2}}= R^{i}_{00j}N^{j}
 +\frac{1}{2}\left(R^{i}_{0 0 j ;
k}-R^{i}_{j k 0 ; 0}\right)N^{j}N^{k} + 2R^{i}_{j 0
k}\frac{dN^{j}}{dt} N^{k} \label{deviation3a}. \ena Clearly, the
last term in both equations is common, and it resembles the
magnetic part of the electromagnetic Lorentz force. However, as
was shown above, for particles which do not have large unperturbed
velocities and are (on average) at rest in the local inertial
frame, this term is quadratic in $h$ and should be neglected. At
the same time, the second term in Eq. (\ref{deviation3a}) (which
we provisionally call ``magnetic") is linear in $h$ and cannot be
neglected. Regardless of terminology, the correct results
exhibited in Eq. (\ref{barxyz}) can only be obtained if one
proceeds with Eq. (\ref{deviation3a}) and not with Eq.
(\ref{deviationmashhoon}).


\section{\label{sec:distances} Variation of the distance between test masses}

We have used a local inertial frame and completely specified the
time-dependent positions of test particles acted upon by a
gravitational wave. As was already mentioned, this description is
as close as possible to the description of laboratory physics.
Having answered all the questions with regard to particles's
positions, we can now discuss the variation of distances between
them. We are mostly interested in the distance between the central
particle and the particle located, on average, at some position
$(l_1, l_2, l_3)$. This is a model for the central mirror and the
end-mirror placed in one of the arms of an interferometer. We will
later use these results for the derivation of the response
function of the laser interferometer.

In the local inertial frame, metric tensor (\ref{limetr}) has the
Minkowski values up to small terms of order of
$h(\frac{l}{\lambda})^2$. The Euclidian expression \bea d(t) =
\sqrt{{\overline{x}^1}^2+{\overline{x}^2}^2+{\overline{x}^3}^2}
+O\left(hl(l/ \lambda)^2\right), \ena gives the distance between
particles, which is accurate up to terms of the order of $hl$ and
$hl\frac{l}{\lambda}$ inclusive. Obviously, we neglect the terms
quadratic in $h$. Denoting ${\bar{x}^i}= l_i + \Delta
{\bar{x}^i}$, one gets \bea d(t) \approx l + \frac{1}{l}\left(l_1
\Delta{\bar{x}^1} + l_2 \Delta{\bar{x}^2} + l_3
\Delta{\bar{x}^3}\right). \label{vard} \ena Using the
time-dependent positions (\ref{barxyz}), we obtain the distance
$d(t)$ with the required accuracy: \bea \fl d(t)= & l +
\frac{1}{2l}\left[h_{+}({l_1}^2-{l_2}^2)\sin{(\omega
t+\psi)}-2h_{\times}{l_1}{l_2}\cos{(\omega t+\psi)}\right] \nonumber \\
\fl & +\frac{1}{4l}kl_3\left[h_{+}({l_1}^2-{l_2}^2)\cos{(\omega
t+\psi)}+2h_{\times}{l_1}{l_2}\sin{(\omega t+\psi)}\right]+
O\left(hl(l/ \lambda)^2\right). \nonumber \\ \fl &
\label{deltar12} \ena Clearly, the first correction to $l$ is due
to the ``electric" contribution, whereas the second correction to
$l$ is due to the ``magnetic" contribution.

According to Eqs. (\ref{barxyz}), the ``magnetic" component of
motion is present even if the mean position of the second mass is
such that $l_3=0$. However, this motion is in the direction
orthogonal to the line joining the masses and therefore it does
not lead to a (first order in terms of $h$) change of distance
between them. This fact is reflected in Eq. (\ref{deltar12}) in
the form of disappearance of the ``magnetic" contribution to the
distance when $l_3 =0$. In other words, ``magnetic" contribution
to the distance is present only if the interferometer's arm is not
orthogonal to the wave's propagation.

The distance (\ref{deltar12}) was calculated in the local inertial
frame. It was assumed that $l/\lambda \ll 1$. It is important to
show that the approximate expression (\ref{deltar12}) follows also
from exact definitions of distance. One of them is based on the
measurement of time that it takes for a light ray to travel from
one free particle to another and back. [This is a part of a more
general problem of finding light-like geodesics in the presence of
a weak gravitational wave \cite{mtw},\cite{gr74},
\cite{estabrook}.] This definition is applicable regardless of the
relationship between $l$ and $\lambda$ and does not require the
introduction of a local inertial frame. If a photon is sent out
from the first particle-mirror at the moment of time $t_0$, gets
reflected off the second particle-mirror, and then returns back to
the first particle-mirror at $t_2$, the proper distance $d(t)$
between the mirrors at time $t$ is defined as \bea
d\left(t\right)=c\frac{t_2-t_0}{2}, \label{exdist} \ena where
$t=(t_0+t_2)/2$ is the mean time between the departure of the
photon and its arrival back.

In the field of the gravitational wave (\ref{metric}), the light
rays, $ds^2 =0$, are described by the equation \bea c^2dt^2=
(1+a)d{x^1}^2+(1-a)d{x^2}^2-2b~d{x^1}d{x^2}+d{x^3}^{2}.
\label{lightray}\ena Let the (unperturbed) outgoing light ray be
parameterized as \bea x^0= ct_0 +l\tau, ~~~x^1=l_1\tau,~~~ x^2
=l_2 \tau, ~~~x^3 =l_3 \tau,\ena where the parameter $\tau$
changes from 0 to 1. Then, according to Eq. (\ref{lightray}), we
have along the ray: \bea \label{light2} cdt = l\left[1+
\frac{1}{2} a(\tau) \frac{l_1^2 -l_2^2}{l^2} - b(\tau) \frac{l_1
l_2}{l^2} \right] d \tau. \ena Integrating both parts of this
equation, we can find the time $t$ of arrival of the photon to the
second particle. The calculated time includes the g.w. corrections
proportional to $h_{+}$ and $h_{\times}$. Similarly, the
(unperturbed) reflected light ray can be parameterized as \bea x^0
= ct +l\tau, ~~~x^1=l_1- l_1 \tau, ~~~x^2=l_2-l_2
\tau,~~~x^3=l_3-l_3 \tau, \ena where $\tau$ is again changing from
0 to 1. A similar integration of Eq. (\ref{light2}) allows us to
find the time $t_2$ of arrival of the photon back to the first
particle, including the g.w. corrections.

Combining two pieces of the light travel time, we derive the exact
formula for the distance: \bea
 d(t) = l+ \frac{1}{2l} \left[ h_{+} \frac{l_1^2-l_2^2}{2} {\Phi}_{c} +h_{\times} l_1 l_2
\Phi_s \right],\label{dist} \ena where \bea \fl \Phi_{c} = & -
\frac{1}{k(l+l_3)} \left[\cos{(\omega t + \psi +kl_3)} -
\cos{(\omega t +\psi -kl)} \right] \nonumber \\ \fl & +
\frac{1}{k(l-l_3)} \left[ \cos{(\omega t + \psi + kl_3)} -
\cos{(\omega t +\psi +kl)} \right],  \ena and $\Phi_s$ is obtained
from $\Phi_c$ by the replacement of all $\cos$-functions with
$\sin$-functions of the same arguments. Exactly the same formula
(\ref{dist}) follows also from the direct joining of the perturbed
light-like geodesic lines, derived in reference \cite{gr74}.

Formula (\ref{deltar12}) is sufficient for ground-based
interferometers, for which the condition $l \ll \lambda$ is
usually satisfied. Formula (\ref{dist}) is appropriate for
space-based interferometers for which the above condition is not
satisfied in the higher-frequency portion of the sensitivity band
(see, for example, \cite{thorne}, \cite{armstrongtinto},
\cite{cornish}, \cite{lisa}). However, it is important that
formula (\ref{deltar12}), including its ``magnetic" terms, follows
also from exact definition (\ref{dist}), when the appropriate
approximation is taken. Assuming that $k(l+l_3) \ll 1$ and
$k(l-l_3) \ll 1$ and retaining only the first two terms in the
expansion of $\Phi_c$ and $\Phi_s$, one derives Eq.
(\ref{deltar12}) from Eq. (\ref{dist}). As expected, the
``magnetic" contribution to the distance is a universal
phenomenon. It is most easily identified and interpreted in the
local inertial frame, but conclusions about the distance do not
depend on the introduction of this frame.

It is known that there is no unique definition of spatial distance
in curved space-time. We have considered the definition based on
measuring the round-trip time of a light ray. One more definition
is based on measuring the length of a spatial geodesic line
joining the particles at a fixed moment of time. It can be shown
that this definition leads to the formula \bea
 d(t) = l+ \frac{1}{2l} \left[ h_{+} \frac{l_1^2-l_2^2}{2} {\tilde{\Phi}}_{c} +h_{\times} l_1 l_2
\tilde{\Phi}_s \right],\label{spatialgeodesicdistance}\ena where
\bea  \tilde{\Phi}_{c} = & - \frac{2}{kl_3} \left[\cos{(\omega t +
\psi +kl_3)} - \cos{(\omega t +\psi )} \right], \ena and
$\tilde{\Phi}_s$ is obtained from $\tilde{\Phi}_c$ by the
replacement of all $\cos$-functions with $\sin$-functions of the
same arguments. In general, the distance
(\ref{spatialgeodesicdistance}) differs from the distance
(\ref{dist}). However, they do coincide in the first order
approximation in terms of small parameter $kl$. This can be shown
by expanding Eq. (\ref{spatialgeodesicdistance}) in powers of $kl$
and retaining the linear terms. It is satisfying that the exact
definitions (\ref{dist}) and (\ref{spatialgeodesicdistance}) lead
precisely to the Euclidean result (\ref{deltar12}) in the
appropriate approximation.


\section{\label{sec:interferometer}Response of an interferometer to the incoming plane gravitational wave}

Laser interferometer measures the difference of distances
travelled by light in two arms. We describe interferometer in the
local inertial frame, with the origin of the frame at the corner
mirror. The unperturbed coordinates of the end-mirrors are given
by $(l^{(\textsl{a})}_{1},~l^{(\textsl{a})}_{2},~
l^{(\textsl{a})}_{3})$, where $\textsl{a}=1,2 $ labels the arms.
We consider an interferometer whose unperturbed arms have equal
lengths,
$l=\sqrt{{l^{(\textsl{a})}_{1}}^2+{l^{(\textsl{a})}_{2}}^2+
{l^{(\textsl{a})}_{3}}^2}$, and the arms are orthogonal to each
other. A detailed introduction to interferometers in
gravitational-wave research can be found in \cite{kt},
\cite{saulson}, \cite{thorne}.

It was shown in the previous section that the distance variation
in one of the arms is given by Eq. (\ref{deltar12}). Then, the
response of a 2-arm interferometer is given by \bea \Delta d(t)  =
d(t)^{(1)} - d(t)^{(2)}. \label{dr} \ena The derivation of formula
(\ref{deltar12}) is based on the coordinate system adjusted to the
gravitational wave. Specifically, it is assumed that the $x^3$
axis is the direction from which the incoming plane wave
propagates, while the $x^1, x^2$ directions are defining the
principal axes of the wave, see Eq. (\ref{metric}). Then, the
response function (\ref{dr}) is characterized by six free
parameters $(l^{(\textsl{a})}_{1},~l^{(\textsl{a})}_{2},~
l^{(\textsl{a})}_{3})$. Among these six parameters only four are
independent (one of which is $l$), because the arms have equal
unperturbed lengths $l$ and are orthogonal to each other.

\begin{figure}
\begin{center}
\includegraphics{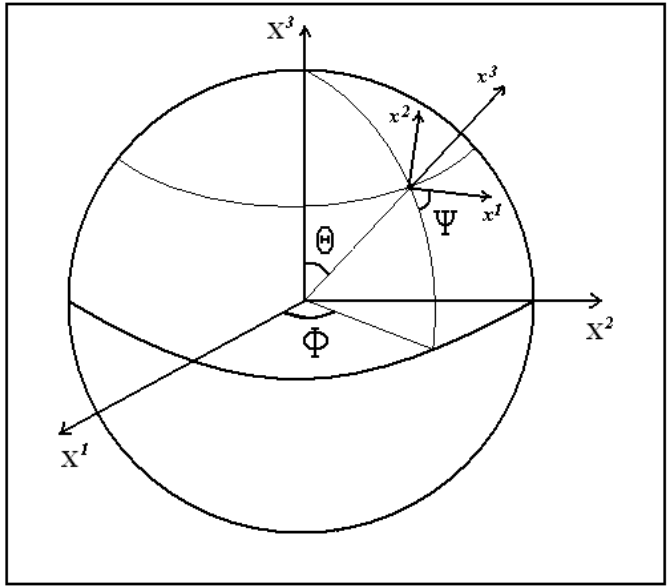}  \end{center}
\caption{\label{fig:sphere} The relationship between the
coordinate system $({x^1},{x^2},{x^3})$ adjusted to the wave and
the coordinate system $({X^1},{X^2},{X^3})$ adjusted to the
interferometer. The two arms of the interferometer lie along the
axes $X^1$ and $X^2$.} \end{figure}

When it comes to the observer, it is more convenient to associate
coordinate system with the interferometer's arms, rather than with
one particular wave. Let the observer's coordinate system
$({X^1},{X^2},{X^3})$ be chosen in such a way that the arms are
located along the $X^1$ and $X^2$ directions. Then, the response
function is characterized by $l$ and three angles : $\Theta$,
$\Phi$ and $\Psi$. The angles $\Theta, \Phi$ describe the
direction of a particular incoming wave, and the third angle
$\Psi$ describes the orientation of the principal axes of the wave
with respect to the observer's meridian (see, for example,
\cite{kt}, \cite{saulson}, \cite{larson}). The relationship
between these two coordinate systems is shown in
Fig.~\ref{fig:sphere} and is described in detail in
\ref{app:coordtransf}.

The parameters $(l^{(\textsl{a})}_{1},~l^{(\textsl{a})}_{2},~
l^{(\textsl{a})}_{3})$ are expressible in terms of the parameters
$\Theta, \Phi, \Psi$ according to the relationships (see
\ref{app:coordtransf}): \bea\begin{array}{l} l^{(1)}_{1}=l(\cos{\Phi}\cos{\Theta}\cos{\Psi}-\sin{\Phi}\sin{\Psi}),\\
l^{(1)}_{2}=l(-\sin{\Phi}\cos{\Psi}-\cos{\Phi}\cos{\Theta}\sin{\Psi}),\\
l^{(1)}_{3}=l(\cos{\Phi}\sin{\Theta}), \end{array} \label{l1}\ena and
\bea\begin{array}{l} l^{(2)}_{1}=l(\sin{\Phi}\cos{\Theta}\cos{\Psi}+\cos{\Phi}\sin{\Psi}),\\
l^{(2)}_{2}=l(\cos{\Phi}\cos{\Psi}-\sin{\Phi}\cos{\Theta}\sin{\Psi}),\\
l^{(2)}_{3}=l(\sin{\Phi}\sin{\Theta}). \end{array} \label{l2} \ena

Using (\ref{l1}) and (\ref{l2}) in (\ref{dr}), the response
function can be written in the form \bea \fl \Delta d(t) = l &
\left[ ~ h_{+} \left\{F_{+}^{E} (\Theta,\Phi,\Psi) \sin (\omega t
+ \psi) + (kl) F_{+}^{H} (\Theta,\Phi,\Psi) \cos (\omega t
+\psi)\right\} \right. \nonumber \\ \fl & \left. +h_{\times}
\left\{F_{\times}^{E} (\Theta,\Phi,\Psi) \cos (\omega t + \psi) +
(kl) F_{\times}^{H} (\Theta,\Phi,\Psi) \sin (\omega t
+\psi)\right\} \right], \label{dr2} \ena where the ``electric"
($E$) components of $F_{+,\times}(\Theta,\Phi,\Psi)$ are given by
\bea \fl F_{+}^{E}(\Theta,\Phi,\Psi) & = & \cos{2\Phi}
\left(\frac{1+\cos^2{\Theta}}{2}\right)\cos{2\Psi}-\sin{2\Phi}
\cos{\Theta}\sin{2\Psi},
\label{electra} \\
 \fl F_{\times}^{E}(\Theta,\Phi,\Psi) & = - &
\cos{2\Phi}
\left(\frac{1+\cos^2{\Theta}}{2}\right)\sin{2\Psi}-\sin{2\Phi}
\cos{\Theta}\cos{2\Psi},\label{electrb} \ena and the ``magnetic"
($H$) components of $F_{+,\times}(\Theta,\Phi,\Psi)$ are given by
\bea \fl F_{+}^{H}(\Theta,\Phi,\Psi) = &
\frac{1}{4}\sin{\Theta}\left[
\left(\cos^2{\Theta}+\sin{2\Phi}\left(\frac{1+\cos^2{\Theta}}{2}\right)
\right)\left(\cos{\Phi}-\sin{\Phi}\right)\cos{2\Psi}
\right.\nonumber \\ \fl  & \left.
-\frac{}{}\sin{2\Phi}\left(\cos{\Phi}+\sin{\Phi}\right)\cos{\Theta}
\sin{2\Psi} \right],\label{magneta}\\ \fl
F_{\times}^{H}(\Theta,\Phi,\Psi) = & \frac{1}{4}
\sin{\Theta}\left[
\left(\cos^2{\Theta}+\sin{2\Phi}\left(\frac{1+\cos^2{\Theta}}{2}\right)
\right)\left(\cos{\Phi}-\sin{\Phi}\right)\sin{2\Psi}
\right.\nonumber \\ \fl & \left.
+\frac{}{}\sin{2\Phi}\left(\cos{\Phi}+\sin{\Phi}\right)\cos{\Theta}
\cos{2\Psi} \right]. \label{magnetb} \ena Our response function is
more accurate than the previously derived expressions \cite{kt},
\cite{saulson}, \cite{larson}, \cite{tinto}, because our Eq.
(\ref{dr2}) includes the ``magnetic" contribution. In equation
(\ref{dr2}) it is given by the terms proportional to the factor
$(kl)$.

In general, the response function (\ref{dr2}) contains two
independent polarization amplitudes, $h_{+}$ and $h_{\times}$. To
simplify the analysis of Eq. (\ref{dr2}), we will separately
consider circularly polarized ($h_{+} = \pm h_{\times}$), and
linearly polarized ($h_{\times} = 0$ or $h_{+} = 0 $), waves. We
start with the analysis of the response function as a function of
time, assuming that a circularly polarized or a linearly polarized
wave arrives from a fixed direction $\Theta, \Phi$ on the sky.

\begin{figure}
\begin{center}
\includegraphics{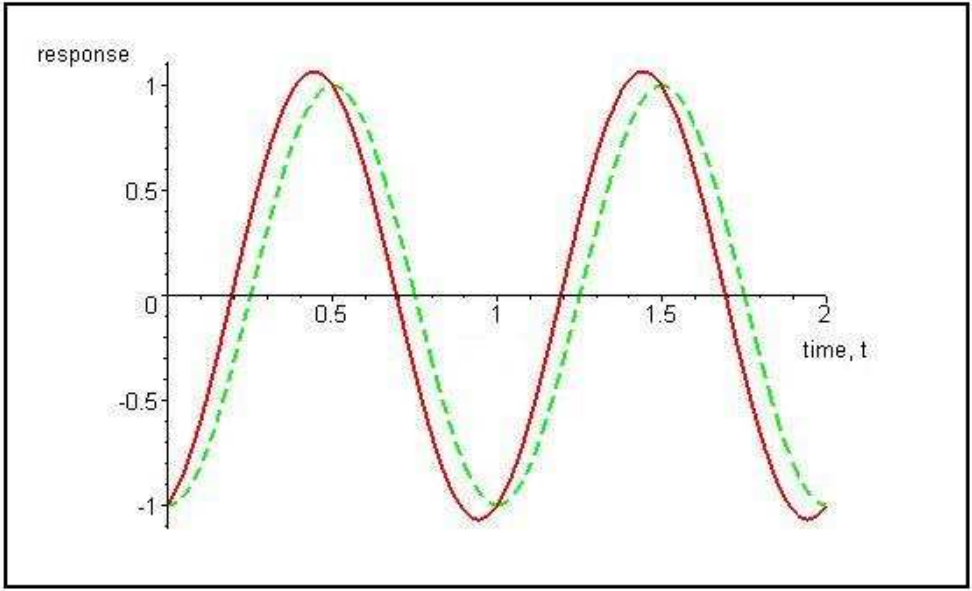}  \end{center}
\caption{\label{fig:timeresp} A typical response of an
interferometer, as a function of time, to the monochromatic
circularly polarized gravitational wave coming from a fixed
direction on the sky. The solid line shows the total response,
while the dashed line is purely ``electric" part.} \end{figure}

In the case of a circularly polarized wave ($h_+ = \pm
h_{\times}$) we obtain \bea \fl \Delta d(t) =
l~h_{+}~{\left[{F^{E}_{+}}^{2}+{F^{E}_{\times}}^{2}
\right]}^{\frac{1}{2}} \left[1 \pm (kl)\left(\frac{F^{E}_{+}
F^{H}_{\times}+F^{E}_{\times} F^{H}_{+}}
{{F^{E}_{+}}^{2}+{F^{E}_{\times}}^{2}}\right)\right]
\sin{\left(\omega t+\psi \pm \Delta\psi\right)}, \label{33} \ena
where the phase shift $\Delta\psi$ is given by \bea
\tan{\left(\Delta\psi\right)} =
\frac{F^{E}_{\times}}{F^{E}_{+}}\left[1\pm (kl)
\left(\frac{F^{E}_{+} F^{H}_{+}-F^{E}_{\times} F^{H}_{\times}}
{F^{E}_{+} F^{E}_{\times}}\right)\right]. \ena Clearly, the
inclusion of ``magnetic" terms changes the amplitude and the phase
of $\Delta d(t)$. As an illustration, we show in
Fig.~\ref{fig:timeresp} the response of an interferometer, as a
function of time, to a circularly polarized wave $h_+ =
h_{\times}$, coming from the direction $\Theta=2\pi/3, \Phi=\pi/3$
(we have also set $\Psi=0$ to fix the phase of the response
function; the amplitude of the response function does not depend
on $\Psi$). The dashed curve shows the ``electric" response alone,
while the solid curve shows the total response, including the
``magnetic" part. We have taken $l/\lambda =0.1$.

The response of an interferometer to any elliptically polarized
wave is qualitatively similar to Fig. \ref{fig:timeresp}, that is,
in general, the ``magnetic" component contributes, both, to the
amplitude and to the phase of $\Delta d(t)$. However, ``magnetic"
contribution to the amplitude may be very small for linearly
polarized waves. If one puts $h_{\times}=0$ or $h_{+}=0$ in Eq.
(\ref{dr2}), one finds that the correction to the amplitude
$\Delta d(t)$ is quadratic in $(kl)^2$ (and, hence, can be
neglected) with the exception of directions on the sky where
$F_+^E$ or $F_{\times}^E$ vanish. Specifically, for the linearly
polarized wave $h_{\times}=0$, we obtain \bea \Delta d(t) =
l~h_{+}~ \sqrt{{F_{+}^{E}}^2+(kl)^2{F_{+}^{H}}^2}\sin{(\omega
t+\psi+\Delta\psi)}, \label{47}\ena where the phase shift
$\Delta\psi$ is given by \bea \tan{\left(\Delta\psi\right)}=
kl\left(\frac{F_{+}^{H}}{F_{+}^{E}}\right). \ena (The case when
$h_{+}=0$ is given by the above expressions in which all the
``plus" indices are replaced by ``cross" indices, and the
sin-function in expression (\ref{47}) is replaced by a
cos-function.)

\begin{figure}
\begin{center}
\includegraphics{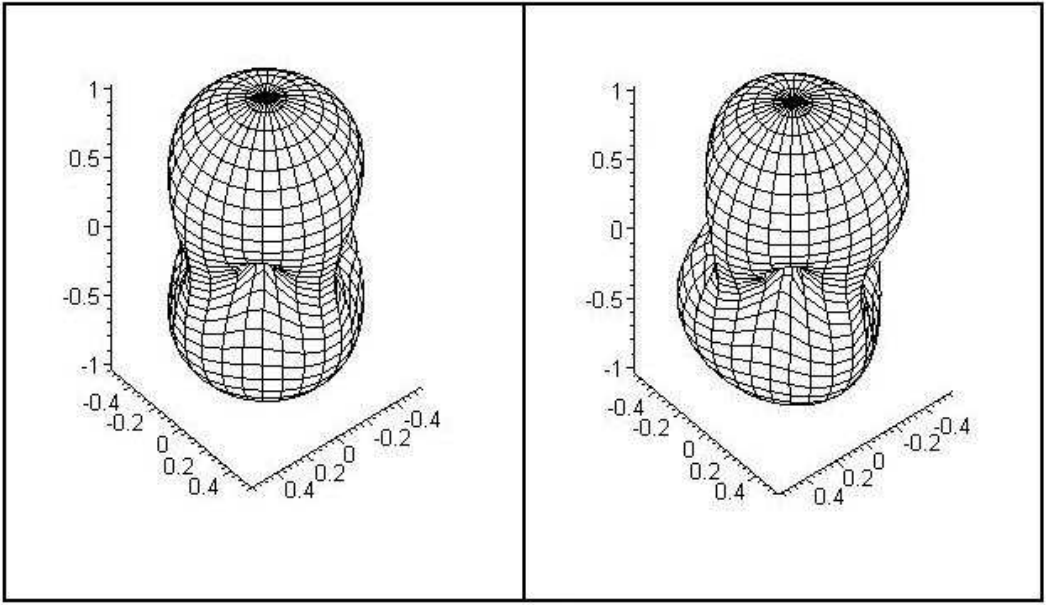}  \end{center}
\caption{\label{fig:skyresp} The amplitude of the interferometer's
response to circularly polarized waves. The graphs are normalized
in such a way that the amplitude is equal 1 for $\Theta =0$. The
left figure ignores the ``magnetic" effect, whereas the right
figure shows the total response.} \end{figure}

We now turn to the amplitude of $\Delta d(t)$ as a function of the
angles $\Theta, \Phi$. It is usually called the beam pattern. In
general, the amplitude of $\Delta d(t)$ depends also on $\Psi$,
but we will consider, for simplicity, circularly polarized waves,
in which case the parameter $\Psi$ does not participate in the
amplitude of $\Delta d(t)$. We consider circularly polarized
waves, with one and the same amplitude $h_R$, coming from
arbitrary directions on the sky. The beam pattern is shown in
Fig.~\ref{fig:skyresp}. The left figure shows the purely
``electric" contribution, whereas the right figure shows the total
beam pattern, with the ``magnetic" contribution included. We have
taken $l/\lambda = 0.1$. It is clearly seen that the ``magnetic"
component breaks the characteristic quadrupole symmetry of the
``electric" beam pattern (more detail - in the next Section).

\section{\label{sec:spinfunction}Response function in terms of spin-weighted spherical harmonics}

The response function (\ref{dr2}) allows an elegant invariant
representation in terms of the spin-weighted spherical harmonics.
This description is also helpful for physical interpretation of
the response function.

First, let us introduce the amplitudes of circularly polarized
gravitational waves, \bea \label{hrhl} h_R= \frac{1}{2}
\left(h_{+} + h_{\times} \right), ~~~~ h_L= \frac{1}{2}
\left(h_{+} - h_{\times} \right), \ena and the complex functions
\bea \label{frfl} \begin{array}{l} F_{R}=\left(F_{+}^E
+(kl)F_{\times}^H\right)+i \left(F_{\times}^E+ (kl)F_{+}^H
\right), \\ \\ F_{L}=\left(F_{+}^E -(kl)F_{\times}^H\right)+i
\left(-F_{\times}^E+ (kl)F_{+}^H \right). \end{array} \ena

In terms of the introduced notations, the response function
(\ref{dr2}) can be identically rewritten as \bea \fl \Delta d(t) =
l \left(-\frac{i}{2}\right) \left[ e^{i(\omega t+\psi)} \left(h_R
F_R +h_L F_L \right) - e^{-i(\omega t+\psi)} \left(h_R F_R^{*}
+h_L F_L^{*} \right)\right].\label{dr3} \ena Clearly, the $\Delta
d(t)$ has been separated in two components associated with the
left (L) and right (R) polarization states of the gravitational
wave. Only the $L(R)$ component of the response function is
present, when the $R(L)$ g.w. amplitude is set to zero.

Using the explicit expressions (\ref{electra}), (\ref{electrb}),
(\ref{magneta}) and (\ref{magnetb}), one can show that the newly
defined functions $F_R, F_L$ can be rearranged to read \bea
\label{frfl2} F_{R}=\widetilde{F}_{R}(\Theta, \Phi)e^{-2i\Psi},
~~~~ F_{L}=\widetilde{F}_{L}(\Theta, \Phi)e^{2i\Psi}, \ena where
$\widetilde{F}_{L}$ and $\widetilde{F}_{R}$ are given by \bea \fl
\widetilde{F}_{L}= &
\frac{1}{4}\left[\left(1+\cos{\Theta}\right)^{2}e^{2i\Phi}+\left(1-\cos{
\Theta}\right)^{2}e^{-2i\Phi}\right] \nonumber
\\ \fl  & +\frac{ikl}{32} \left[ (1-i)\left(1+\cos{\Theta}\right)^{2}\sin{\Theta}e^{3i\Phi}+
(1+i)\left(
3\cos{\Theta}-1\right)\left(\cos{\Theta}+1\right)\sin{\Theta}e^{i\Phi}\right.
\nonumber\\  \fl & \left.
+(1-i)\left(3\cos{\Theta}+1\right)\left(\cos{\Theta}-1\right)\sin{\Theta
}e^{-i\Phi}+
(1+i)\left(1-\cos{\Theta}\right)^{2}\sin{\Theta}e^{-3i\Phi}
\right], \nonumber
\\   \fl & \ena \bea \fl
\widetilde{F}_{R}= &
\frac{1}{4}\left[\left(1-\cos{\Theta}\right)^{2}e^{2i\Phi}+\left(1+\cos{
\Theta}\right)^{2}e^{-2i\Phi}\right]  \nonumber
\\ \fl &
 +\frac{ikl}{32} \left[ (1-i)\left(1-\cos{\Theta}\right)^{2}\sin{\Theta}e^{3i\Phi}+
(1+i)\left(
3\cos{\Theta}+1\right)\left(\cos{\Theta}-1\right)\sin{\Theta}e^{i\Phi}\right.
\nonumber \\ \fl & \left. \
+(1-i)\left(3\cos{\Theta}-1\right)\left(\cos{\Theta}+1\right)\sin{\Theta
}e^{-i\Phi}+
(1+i)\left(1+\cos{\Theta}\right)^{2}\sin{\Theta}e^{-3i\Phi}
\right]. \nonumber \\ \fl & \ena For a given direction $(\Theta,
\Phi)$ on the sky, the functions $F_R$ and $F_L$ transform under
the rotation of $\Psi$, specified by the transformation $\Psi
\rightarrow \Psi+\Psi_{0}$, according to
$\widetilde{F}_{R}\rightarrow \widetilde{F}_{R}e^{- 2i\Psi_{0}}$
and $\widetilde{F}_{L}\rightarrow
\widetilde{F}_{L}e^{2i\Psi_{0}}$. A function $ {
}_{s}f(\Theta,\Phi)$ is said to be a spin-$s$ function, if under
the transformation $\Psi \rightarrow \Psi+\Psi_{0}$ it transforms
like ${ }_{s}f(\Theta,\Phi)\rightarrow { }_{s}f(\Theta,\Phi)e^{-
is\Psi_{0}}$. Therefore, our functions $F_{L}$ and $F_{R}$
represent the $s=-2$ and $s=+2$ functions, respectively.

A scalar function on a 2-sphere can be expanded over a set of
ordinary spherical harmonics $Y^{l}_{m}(\Theta,\Phi)$, which form
a complete and orthonormal basis. These harmonics are not
appropriate for the expansion of spin-weighted functions with $s
\neq 0$. The spin-weighted functions can be expanded over a set of
harmonics called the spin-weighted spherical harmonics $ {
}_{s}{Y_{m}^{l}}  (\Theta,\Phi)$ \cite{goldberg}, \cite{thorne2},
\cite{penrose}, \cite{zaldkam}. The spin-weighted spherical
harmonics satisfy the completeness and othonormality conditions
similar to those of ordinary spherical harmonics. The functions $
{ }_{s}{Y_{m}^{l}}(\Theta, \Phi)$ can be derived from
$Y^{l}_{m}(\Theta, \Phi)$ according to the rules: \bea {}_{s}Y^{l}_{m}= \left\{ \begin{array}{l}
\left[\frac{(l-s)!}{(l+s)!}\right]^{1/2}\eth^{s}Y^{l}_{m} ,\ \ (0 \leq s \leq l) \\ \\
\left(-1\right)^{s}\left[\frac{(l+s)!}{(l-s)!}\right]^{1/2}\bar{\eth}^{-
s}Y^{l}_{m} ,\ \ (0 \geq s \geq -l)
 \end{array}
 \right. ,
\ena where the differential operators $\eth$ and $\bar{\eth}$
[acting on a spin $s$ function $ { }_{s}f(\Theta,\Phi)$] are given
by \bea \begin{array}{l}\eth ~ { }_{s}f(\Theta,\Phi) =
-\sin^{s}{\Theta}
\left(\frac{\partial}{\partial\Theta}+\frac{i}{\sin{\Theta}}\frac{\partial}{\partial\Phi}\right)
\sin^{-s}{\Theta}~ { }_{s}f(\Theta,\Phi), \\ \\
\bar{\eth}~  { }_{s}f(\Theta,\Phi) = -\sin^{-s}{\Theta}
\left(\frac{\partial}{\partial\Theta}-\frac{i}{\sin{\Theta}}\frac{\partial}{\partial\Phi}\right)
\sin^{s}{\Theta}~ { }_{s}f(\Theta,\Phi). \end{array}\ena

Using the spin-weighted spherical harmonics, the functions
$\widetilde{F}_{L}$ and $\widetilde{F}_{R}$ can be expanded as
follows: \bea \fl \widetilde{F}_{L} = & \sqrt{2\pi}\left[
{}_{-2}Y^{2}_{2}(\Theta,\Phi)+
{}_{-2}Y^{2}_{-2}(\Theta,\Phi)\right]\nonumber\\ \fl  &
+ikl\sqrt{\frac{\pi}{46}}\left[ ~ ~ \frac{(1-i)}{\sqrt{3}} {}_{-2}
Y^{3}_{3}(\Theta,\Phi) +\frac{(1+i)}{\sqrt{5}} {}_{-2}
Y^{3}_{1}(\Theta,\Phi) \right. \nonumber\\ \fl &
\qquad\qquad\left. +\frac{(1-i)}{\sqrt{5}} {}_{-2}
Y^{3}_{-1}(\Theta,\Phi)
+\frac{(1+i)}{\sqrt{3}} {}_{-2}Y^{3}_{-3}(\Theta,\Phi) \right],\\
\fl \widetilde{F}_{R}= & \sqrt{2\pi}\left[
{}_{2}Y^{2}_{2}(\Theta,\Phi)+
{}_{2}Y^{2}_{-2}(\Theta,\Phi)\right]\nonumber\\ \fl
  & +ikl\sqrt{\frac{\pi}{46}}\left[ ~ ~ \frac{(1-i)}{\sqrt{3}} {}_{2} Y^{3}_{3}(\Theta,\Phi) +\frac{(1+i)}{\sqrt{5}} {}_{2}
Y^{3}_{1}(\Theta,\Phi) \right. \nonumber\\ \fl & \qquad\qquad
\left. +\frac{(1-i)}{\sqrt{5}} {}_{2}Y^{3}_{-1}(\Theta,\Phi)
+\frac{(1+i)}{\sqrt{3}} {}_{2} Y^{3}_{-3}(\Theta,\Phi) \right].
 \ena  As one could expect, the ``electric" components of
the response function are described by the spin 2 quadrupole (l=2)
terms ${}_{\pm 2}Y^{2}_{m}$, whereas the ``magnetic" components
are described by the spin 2 octupole (l=3) terms ${}_{\pm
2}Y^{3}_{m}$. The ``magnetic" components can be viewed as the
higher-order terms in the multipole expansion of the response
function.

\section{\label{sec:discussion}Astrophysical example}

The ``magnetic" component in the motion of free masses will have
serious practical implications for the current gravitational wave
observations. For example, the LIGO interferometers have the arm
length of $l=4km$ and are most sensitive to g.w. in the interval
of frequencies between $30Hz$ and $10^{4}Hz$. This means that the
``magnetic" component, whose relative contribution is of the order
of $kl$, may provide a correction at the level of $5 \%$ to the
response function in the frequency region of $600 Hz$, and up to
$10 \%$ in the frequency region of $1200 Hz$. This contribution
may significantly affect the determination of the source's
parameters.

The high frequency region, $600Hz-1200Hz$, will be extensively
studied with the help of the ``narrow-band tuning" of advanced
interferometers \cite{cutlthorne}. The signal recycling technique
allows reshaping of the sensitivity curve in such a way that it
reduces noise within a chosen narrow band, while weakening
sensitivity in other frequency regions. This region of high
frequencies is populated by various periodic and quasi-periodic
astrophysical sources, such as neutron stars in low-mass X-ray
binaries, including Sco X-1, slightly deformed rotating neutron
stars, neutron star and black hole binaries in the last moments of
their inspiral. For a more detailed list of high frequency sources
and the prospects of their detection see \cite{cutlthorne}.

To better understand the role of the ``magnetic" components in
estimation of the g.w. source parameters, we shall briefly
consider an idealized example of a compact binary system. Let the
binary consist of objects of equal mass $M$ orbiting each other in
a circular orbit of size $2r$. The distance of the binary to the
observer is $R$. For simplicity, we assume that the orbital plane
is orthogonal to the line of sight defined by $\Theta, \Phi$.
Then, the gravitational wave at Earth has the form of Eq.
(\ref{ab}) with the amplitudes given by \cite{saulson},
\cite{glpps}: \bea
\begin{array}{l}
h_{+}= h_{\times}=h_R =\frac{32\pi^{2}G}{Rc^4} Mr^{2}\omega^{2},
\end{array} \ena where the g.w. angular frequency
$\omega=\sqrt{\frac{GM}{r^3}}$ is twice the orbital frequency. The
angle $\Psi$ can be taken as $\Psi =0$, and the phase $\psi$ is
the angle between the observer's meridian and the line joining the
components of the binary at some initial moment of time.

The response of the interferometer to the incoming wave is
determined by Eq. (\ref{dr2}): \bea \fl \Delta d(t)  & =  lh_R ~
\left[\left\{ \cos{2\Phi}\left(\frac{1+ \cos^2{\Theta}}{2}\right)
-\frac{1}{4}\frac{\omega l}{c}
\sin{2\Phi}\left(\cos{\Phi}+\sin{\Phi}\right)\cos{\Theta}\sin{\Theta}\right\}\right.
\nonumber\\ \fl &  \left.\times\sin \left(\omega t+\psi \right)
\right. \nonumber\\ \fl &
 - \left\{\sin{2\Phi}\cos{\Theta} -\frac{1}{4}\frac{\omega l}{c}
 \left(\cos^2{\Theta}+\sin{2\Phi}\left(\frac{1+\cos^2{\Theta}}{2}
\right)\right)
\left(\cos{\Phi}-\sin{\Phi}\right)\sin{\Theta}\right\} \nonumber\\
\fl & \left. \frac{}{}\times \cos{\left(\omega t+\psi\right)}
\right]. \label{examp} \ena The terms with $\omega l$ represent
``magnetic" contribution. If the observed $\Delta d(t)$ is
incorrectly interpreted as the ``electric" contribution only, then
the parameters of source (for example, its mass $2M$) would be
estimated from the relationship: \bea \fl \Delta d(t) =l~h_R~
\left[\cos{2\Phi}\left(\frac{1+\cos^2{\Theta}}{2}\right)
\sin{\left(\omega t+\psi \right)} - \sin{2\Phi}\cos{\Theta}
\cos{\left(\omega t+\psi\right)} \right]. \label{examp2} \ena
Clearly, this would have resulted in a significant error in the
estimated $M$ (of the order $\frac{2\pi l}{\lambda}$). The correct
procedure is the comparison of the observed $\Delta d(t)$ with the
full response function, consisting of ``electric" and ``magnetic"
contributions. A concrete value of the error depends on the
direction $\Theta, \Phi$.

\section{\label{sec:conclusion}Conclusions}

In this paper we have considered the motion of free test particles
in the field of a gravitational wave. We have shown that this
motion is similar to the motion of charged test particles in the
field of an electromagnetic wave. Using different methods we have
demonstrated the presence and importance of what we call the
``magnetic" components of motion. Regardless of interpretation and
terminology, these terms contribute to the variation of distance
between the interferometer's mirrors and, hence, they contribute
to the total response function of the interferometer. The
``magnetic" contribution must be taken into account in advanced
data analysis programs.



\appendix

\section{\label{app:coordtransf1}The principal axes for a monochromatic gravitational wave}

In general, the metric components are given by Eq. (\ref{abg}),
where $\psi_{+}$ and $\psi_{\times}$ are arbitrary independent
constants. However, it is always possible to do a rotation in the
$(x^1, x^2)$ plane, such that in the new coordinate system
$\tilde{x}^1, \tilde{x}^2$ the difference between $\psi_{\times}$
and $\psi_{+}$ is exactly $\frac{\pi}{2}$. The required
transformation \bea \begin{array}{l} \tilde{x}^1=~
x^1\cos{\zeta}+x^2\sin{\zeta},~~~~~
\tilde{x}^2=-x^1\sin{\zeta}+x^2\cos{\zeta},
\end{array}
\ena has the angle $\zeta$ such that \bea \fl
\zeta=\frac{1}{2}\arctan{\left[
\left(\frac{h_{\times}^2-h_{+}^2}{2h_{+}h_{\times}\cos{\left(\psi_{+}-\
psi_{\times}\right)}}\right)\pm
\sqrt{{\left(\frac{h_{\times}^2-h_{+}^2}{2h_{+}h_{\times}\cos{\left(\psi
_{+}-\psi_{\times}\right)}}\right)}^{2}+1}\right]}, \ena The
rotation angle $\zeta$ is well defined, except for a circularly
polarized wave, in which case the difference between
$\psi_{\times}$ and $\psi_{+}$ is always $\frac{\pi}{2}$. In this
new coordinate system $(\tilde{x}^1,\tilde{x}^2)$ the metric
components take the form \bea \begin{array}{l} \tilde{a} =
\tilde{h}_{+}\sin{\left(k({x^0}+{x^3})+\psi\right)},~~~~ \tilde{b}
= \tilde{h}_{\times}\cos{\left(k({x^0}+{x^3})+\psi\right)},
\end{array}
\label{principalaxes} \ena where the amplitudes $\tilde{h}_{+}$
and $\tilde{h}_{\times}$ are given by \bea \fl \begin{array}{l}
\tilde{h}_{+}= {\left(h_{+}^2\cos^{2}{2\zeta} +2h_{+}h_{\times}
\sin{2\zeta}\cos{2\zeta}\cos{\left(\psi_{+} -\psi_{\times}\right)+
h_{\times}^2\sin^{2}{2\zeta}}\right)}^{\frac{1}{2}},
\\ \\
\tilde{h}_{\times}= {\left(h_{\times}^2\cos^{2}{2\zeta}
-2h_{+}h_{\times}\sin{2\zeta}\cos{2\zeta}\cos{\left(\psi_{+}
-\psi_{\times}\right)+h_{+}^2\sin^{2}{2\zeta}}\right)}^{\frac{1}{2}},
\end{array}
\ena and the phase $\psi$ is given by \bea
\psi=\arctan{\left[\frac{h_{+}\sin{\psi_{+}}+h_{\times}\sin{\psi_{\times
}}\tan{2\zeta}}
{h_{+}\cos{\psi_{+}}+h_{\times}\cos{\psi_{\times}}\tan{2\zeta}}\right]}.
\ena We call the coordinates $\tilde{x}^1$ and $\tilde{x}^2$, in
which the metric components take the form (\ref{principalaxes}),
the principal axes of the gravitational wave. In the text, we
assume that the above transformation has already been performed.
So, without loss of physical generality, we put
$\psi_{+}=\psi_{\times}- \frac{\pi}{2}= \psi$, and we do not use
the tilde-symbol.

\section{\label{app:coordtransf}Coordinate transformation}

A general gravitational wave (\ref{metric}), (\ref{ab}) is
described in the coordinate system $(x^1, x^2, x^3)$. For the
observer, it is convenient to use the coordinate system $(X^1,
X^2, X^3)$ in which the arms of interferometer are located along
the $X^1$ and $X^2$ axes. In this coordinate system, the incoming
wave is characterized by the direction $\Theta, \Phi$ and the
angle $\Psi$ between the observer's meridian and the wave's
principal axis, as depicted in Fig.~\ref{fig:sphere}. The
relationship between these two coordinate systems is given by the
transformation \bea \left( \begin{array}{c} {X^1}\\{X^2}\\{X^3}
\end{array} \right)
 =
 R
\left(
\begin{array}{c}
{x^1}\\{x^2}\\{x^3}
\end{array}
\right),
 \ena
where the orthogonal rotation matrix $R$ is \bea {\small \fl
\left( \begin{array}{ccccccc}
\cos{\Phi}\cos{\Theta}\cos{\Psi}-\sin{\Phi}\sin{\Psi} &
-\sin{\Phi}\cos{\Psi}-\cos{\Phi}\cos{\Theta}\sin{\Psi} &
\cos{\Phi}\sin{\Theta}\\ \\
\sin{\Phi}\cos{\Theta}\cos{\Psi}+\cos{\Phi}\sin{\Psi} &
\cos{\Phi}\cos{\Psi}-\sin{\Phi}\cos{\Theta}\sin{\Psi} &
\sin{\Phi}\sin{\Theta}\\ \\ -\sin{\Theta}\cos{\Psi} &
\sin{\Theta}\sin{\Psi} & \cos{\Theta} \end{array} \right). }\ena

In the observer's frame, the end-mirrors have the coordinates
$(X^1 =l, X^2=0, X^3=0)$ and $(X^1 =0, X^2=l, X^3=0)$. Therefore,
in the g.w. frame, the coordinates of the end-mirrors are \bea \fl
l^{(1)}_i= R^{-1} ~ l ~ \left( \begin{array}{c} 1\\0\\0
\end{array} \right) =l ~ \left(
\begin{array}{c} \cos{\Phi}\cos{\Theta}\cos{\Psi}-\sin{\Phi}\sin{\Psi} \\ \\
 -\sin{\Phi}\cos{\Psi}-\cos{\Phi}\cos{\Theta}\sin{\Psi} \\ \\ \cos{\Phi}\sin{\Theta} \end{array} \right),  \ena
  and  \bea \fl l^{(2)}_i= R^{-1} ~ l ~ \left( \begin{array}{c} 1\\0\\0 \end{array}
\right) =l ~ \left(
\begin{array}{c} \sin{\Phi}\cos{\Theta}\cos{\Psi}+\cos{\Phi}\sin{\Psi} \\ \\
 \cos{\Phi}\cos{\Psi}-\sin{\Phi}\cos{\Theta}\sin{\Psi}\\ \\ \sin{\Phi}\sin{\Theta} \end{array} \right).
\ena These formulae allow us to parameterize the positions of the
end-mirrors in terms of $l$ and the angles $\Theta, \Phi, \Psi$.


\end{document}